\documentclass [5p, number] {elsarticle}
\usepackage{graphicx}
\usepackage{setspace}
\usepackage{lineno}
\usepackage{amsmath}
\usepackage{amsmath}

\title{High-rate axial-field ionization chamber for particle identification of radioactive beams}
\date{\today}

\author[chem,ceem]{J. Vadas}
\author[chem,ceem]{Varinderjit Singh}
\author[ceem]{G. Visser}
\author[chem]{A. Alexander}
\author[chem,ceem]{S. Hudan}
\author[chem,ceem]{J. Huston}
\author[chem,ceem]{B. B. Wiggins}
\author[ganil]{A. Chbihi}
\author[wmu]{M. Famiano} 
\author[wmu]{M.M. Bischak}
\author[chem,ceem]{R. T. deSouza \corref{cor1}}

\address[chem]{Department of Chemistry, \\ 
Indiana University, 800 E. Kirkwood Ave., Bloomington, Indiana 47405, USA}

\address[ceem]{Center for Exploration of Energy and Matter, \\ 
Indiana University, 2401 Milo B. Sampson Lane, Bloomington, Indiana 47408, USA}

\address[ganil]{GANIL, \\ 
1 Blvd. Henri Becquerel, Caen, 14000, France}

\address[wmu]{Department of Physics, \\ 
Western Michigan University, Kalamazoo, Michigan 49008, USA}

\cortext[cor1]{desouza@indiana.edu}

\begin{document}
%\linenumbers
\begin{abstract}
The design, construction and performance characteristics of a simple axial-field ionization chamber suitable for identifying ions in a radioactive beam are presented. Optimized for 
use with low-energy radioactive beams ($<$ 5 MeV/A) the detector presents only three 0.5 $\mu$m/cm$^2$ foils to the beam in addition to the detector gas. A fast charge sensitive amplifier (CSA) 
integrated into the detector design is also described. Coupling this fast CSA to the axial field ionization chamber produces an output pulse with a risetime of 60-70 ns and a fall time of 100 ns,
making the detector capable of sustaining a relatively high rate. Tests with an $\alpha$ source establish the detector energy resolution as $\sim$8 $\%$ for an energy deposit of $\sim$3.5 MeV.  
The energy resolution with beams of 2.5 and 4.0 MeV/A $^{39}$K ions and the dependence of the energy resolution on beam intensity is measured. 
At an instantaneous rate of 3 x 10$^5$ ions/s
the energy resolution has degraded to 14\% with a pileup of 12\%.
The good energy resolution of
this detector at rates up to 3 x 10$^5$ ions/s makes it an effective tool in the characterization of low-energy radioactive beams.
\end{abstract}

\maketitle

\section{Introduction}

The development of radioactive isotope beams (RIBs) has enabled the investigation of nuclei
away from $\beta$-stability \cite{Liang03,Lemasson09,Penionzhkevich10,Glasmacher98}, 
which is crucial in understanding nucleosynthesis in exotic astrophysical 
environments \cite{Penionzhkevich10}, as well as the structure of exotic 
nuclei \cite{Penionzhkevich10,Glasmacher98}. RIBs can be produced by a variety of techniques 
including projectile fragmentation \cite{Symons79}, ISOL \cite{Ravn79}, and photofission \cite{Oganessian02}. 
As the primary nuclear reaction produces 
a distribution of product nuclei, to provide a useful secondary radioactive beam, it is necessary to select the nuclide 
of interest from this distribution. This separation is typically accomplished through electromagnetic means \cite{Sherrill91}. 
Often, however, this separation does not produce a pure beam, and other reaction products with 
similar mass-to-charge ratios contaminate the beam. 
An effective way to address this challenge is to identify each ion in the beam on a particle-by-particle basis.
A commonly used approach to accomplish this is through measurement of the energy loss ($\Delta$E) 
and time-of-flight (TOF) of each ion \cite{Fukuda13}.

The $\Delta$E information is often obtained from
gas ionization chambers \cite{Fulbright79} used in transmission mode. 
The uniform thickness achieveable with these detectors, as well as their robustness against radiation damage are 
key factors in their utility. 
A common geometry of an ion chamber utilizes an electric field transverse to the incident ion direction \cite{Fowler75}, 
with a Frisch grid used to remove the dependence of the pulse amplitude on position. 
In this geometry of ion chambers, however, the drift time of the electrons in the direction transverse to the beam direction typically 
limits the rate to $\sim$10$^{4}$ ions/s. To overcome this limitation of a slow response time, an axial 
field design can be employed. 
Axial field ionization chambers have been successfully employed as $\Delta$E detectors for heavy-ions since the 1980's 
\cite{Zurmuhle82,Bandyopadhyaya89}. Recently, the high rate capability of this design has been exploited at both high \cite{Kimura05} 
and low energies \cite{Chae14} by 
using multiple tilted electrodes in the beam path to identify reaction products in a radioactive beam. When the incident energy is 
sufficiently high ($\sim$100 MeV/A) thin metallized foils have been utilized for the electrodes. In contrast, at low 
incident energies ($\sim$5 MeV/A) the electrodes consist of wire harps. Both designs involve the insertion of multiple electrode 
planes (15-24) into the beam path, a source of considerable scattering and energy loss for the incident ion. 
In the subsequent sections, we describe the development and characterization 
of a simple high-rate axial-field ionization chamber, designated the Rare Ion Purity Detector (RIPD) which 
inserts a minimal amount of material into the beam path.

\begin{figure}
\begin{center}
\includegraphics[scale=0.333]{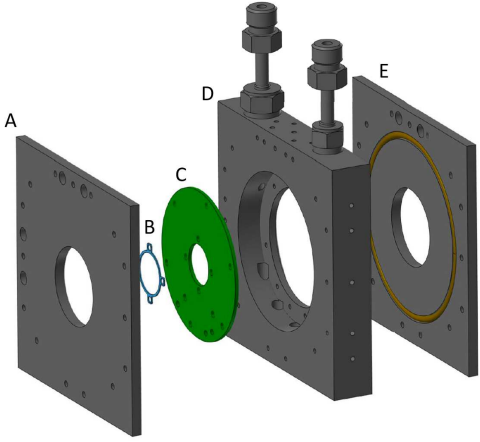}
\caption{
CAD drawing of the Rare Ion Purity Detector in an exploded view. 
A: window frame. B: anode ring. C: PCB. D: RIPD. body E: window frame.
The mylar windows and central anode foil are omitted for clarity.}
\label{fig:RIPD_CAD}
\end{center}
\end{figure}

\section{Physical Configuration}

The principal motivation in developing this detector was investigation of fusion for neutron-rich light ion beams at energies 
near the fusion barrier \cite{Steinbach14,Singh16}. 
Due to the low energy beams utilized in these experiments, particular attention 
was given to the total thickness of the detector in 
order to minimize the beam divergence and energy straggling incurred by inserting the detector into the beam path. Prior experience established that minimizing scattering of the beam was essential. This requirement meant eliminating any wire planes in the beam path.
To implement the simplest axial field geometry while minimizing the electron collection time, a central anode is used with the metallized windows serving as cathodes.

A CAD drawing of RIPD is shown in Fig.~\ref{fig:RIPD_CAD}. The detector is comprised of an aluminum 
body, two stainless steel window plates, and a thin central anode foil. The anode is coupled 
to a charge sensitive amplifier (CSA) housed inside the aluminum body.
The body measures approximately 11 cm x 11 cm transverse to the beam direction and 2.0 cm thick along the beam axis. 
The window plates are 5 mm thick and have a 38 mm diameter opening for the beam to pass through, over which 0.5 $\mu$m 
aluminized mylar is epoxied. These mylar foils serve to contain the gas within the active 
volume and act as cathodes. The window plates are sealed to the body of the detector using O-rings. 
No support wires 
are used with these windows to minimize scattering of the incident beam. Using this geometry, 
it is possible to operate the detector 
at a pressure of 30 torr of CF$_{4}$ for several days without any noticeable degradation 
in the window performance. Repeatedly filling the detector with gas also did not cause a noticeable deterioration 
in the mylar window.
The 0.5 $\mu$m mylar anode foil, which is doubly aluminized, 
is mounted on a 2.0 cm diameter stainless steel ring. This ring is attached to an annular 
printed circuit 
board (PCB) with an inner diameter of 2.0 cm and an outer diameter of 7.4 cm. 
The position of the PCB is chosen so that the distance between the anode foil and each cathode foil is 1 cm.
Holes in the PCB allow gas to flow between the two halves of the detector. This arrangement of the  anode and CSA in close proximity minimizes any additional capacitance 
at the input of the CSA.

\begin{figure*}
\begin{center}
\includegraphics[scale=0.58]{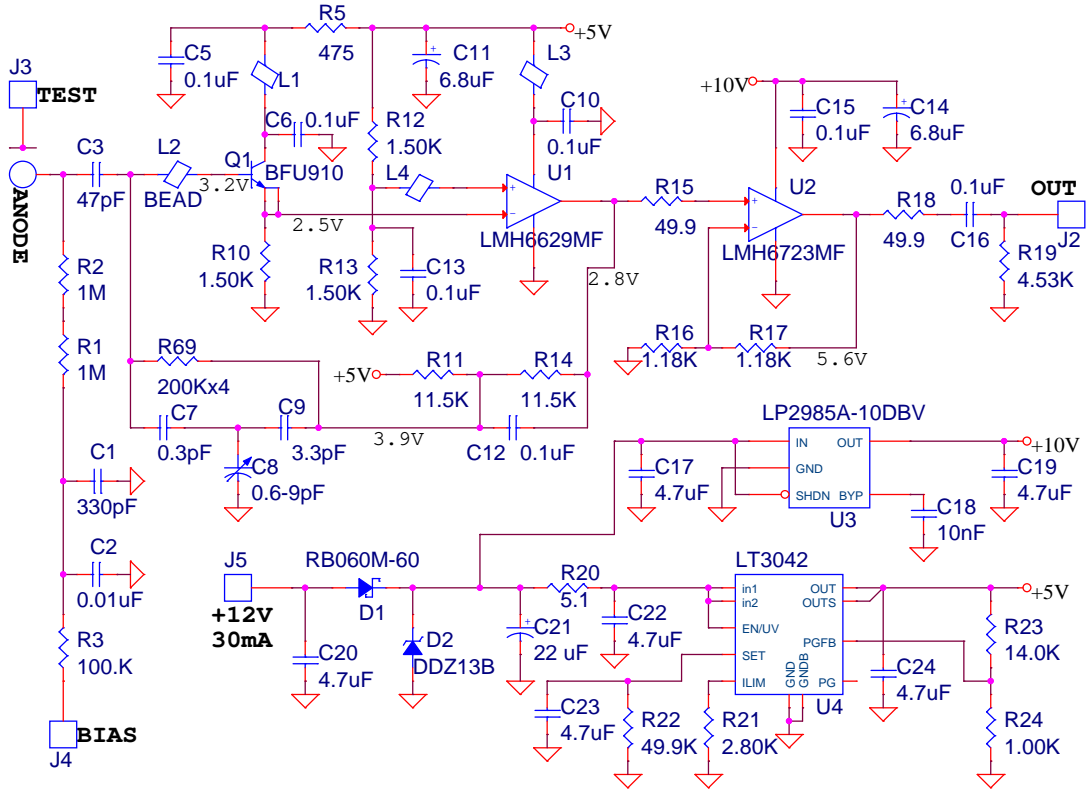}
\caption{
Schematic of the charge sensitive amplifier for RIPD.}
\label{fig:RIPD_CSA}
\end{center}
\end{figure*}

\section{Charge Sensitive Amplifier}

As a key goal in the design of this axial field ionization chamber is its ability to maintain good energy 
resolution while sustaining a high rate 
with minimimum pileup of signals,
it was necessary to develop a fast, low-noise charge sensitive amplifier. 
Ionization of the detector gas induced by a 
beam particle traversing RIPD quasi-instantaneously 
produces an ionization track in the detector. Electrons in this track migrate under the influence of 
the applied electric field and are 
collected at the central anode. 
It should be noted that in contrast to the tilted foil design \cite{Kimura05, Chae14} 
the electric field in RIPD does not move the electrons away from the path of the ionizing beam.
Thus, in comparison to the tilted foil geometry the effects of recombination and screening are expected to 
be larger. This disadvantage is offset by the simplicity of
the present design. 
Carbon tetrafluoride (CF$_4$) was chosen as the detector gas due to its high electron 
drift velocity \cite{Vavra93}. 
Based upon the electron drift velocity for a reduced field of 1 kV cm$^{-1}$ atm$^{-1}$, a rise 
time of 100 ns is anticipated. This 
charge collection time defined one of the necessary characteristics of the CSA. To minimize the 
impact of stray capacitance, the CSA was situated on 
the PCB as close as practically possible to the central anode. The input capacitance of the detector was 
calculated to be 2.25 pF, which was confirmed by measurement. 

The CSA is a new design that is intended to enable high count rates for low capacitance detectors. 
The schematic of the CSA is shown in Fig.\ref{fig:RIPD_CSA}.
The first opamp (​U1) 
provides most of the gain with only a small contribution to the overall noise. However, its bias current, 
input current noise, and input capacitance are too high for direct connection to the anode, so a SiGe microwave 
transistor (Q1) is added to serve as an input buffer. This particular transistor offers very high current gain 
(about 2000) and extremely low input capacitance, and is also very quiet. The signal response of this 
composite amplifier is defined by the feedback network of R69 and C7, 8, and 9. R69 is actually 4 chip 
resistors totaling 800 kohms, which were stacked end to end to minimize stray capacitance. 
The capacitors form a network with an equivalent capacitance adjustable from 0.1 to 0.3 pF. 
The anode board also comprises an output buffer with a gain of 2, two voltage regulators (U4 is remarkably 
low noise), 
level shift, and anode bias circuits. This circuit was realized on the annular FR4 printed circuit 
board on which the RIPD anode was mounted.

\section{Experimental setup}

\begin{figure}
\begin{center}
\includegraphics[scale=0.25]{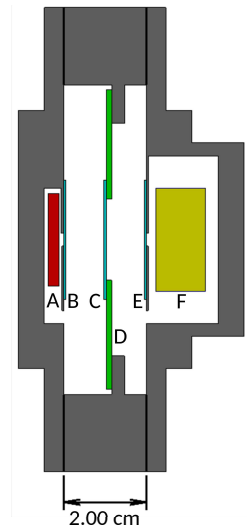}
\caption{
Cross-sectional view of the experimental setup to measure the energy loss of $\alpha$ particles. An $^{241}$Am $\alpha$ source 
was placed just upstream of the entrance foil in a cap that was made common with the active volume. 
Alpha particles that pass through the full length of the detector were then measured in a silicon 
surface barrier detector (SBD) placed just downstream of the exit window in a cap that was also made common with the active volume. 
A: $^{241}$Am $\alpha$ source. B: Entrance cathode foil. C: Anode foil. D: CSA. E: Exit cathode foil. F: SBD.}
\label{fig:RIPD_Setup}
\end{center}
\end{figure}

To characterize the performance of RIPD, the energy resolution for $\alpha$ particles from an 
$^{241}$Am source was measured. 
To test RIPD with $\alpha$ particles, which have a low ionization density,
it is necessary to operate RIPD at gas pressures that exceed the maximum 
pressure sustainable with the thin windows. 
For these tests, the entrance and exit window plates were replaced with flanges, as shown in Fig \ref{fig:RIPD_Setup}. 
The entrance flange allowed an $^{241}$Am source 
to be situated prior to the entrance cathode foil but within the gas volume. Correspondingly, the exit flange allowed a 
silicon surface barrier detector (SBD) to be placed after the exit cathode foil within the gas volume.
Both the $^{241}$Am source and SBD had a 
collimation of 3 mm. Triggering on the SBD signal associated with arrival of $\alpha$ particles 
selected particles that had traversed the entire thickness of RIPD. 
Using this configuration, signals in RIPD could be examined for pressures between 100 and
400 torr. At a pressure of 100 torr the 5.48 MeV $\alpha$ particles deposit just 680 keV in the gas. This energy
deposit increases to $\sim$3.5 MeV at 400 torr. 

%\begin{table}
%\begin{center}
%\begin{tabular}{c c}
%Pressure (torr) & $\Delta$E (MeV) \\
%\hline
%400 & 3.494 \\
%350 & 2.871 \\
%300 & 2.355 \\
%250 & 1.877 \\
%200 & 1.460 \\
%150 & 1.068 \\
%100 & 0.684 \\
%\end{tabular}
\ %caption{
%Energy loss of the $\alpha$ particles through the active volume of RIPD for a given pressure. $\Delta$E values calculated using SRIM \cite{SRIM}.}
%\label{tab:Eloss_table}
%\end{center}
%\end{table}

\section{Signals}

\begin{figure}
\begin{center}
\includegraphics[scale=0.4]{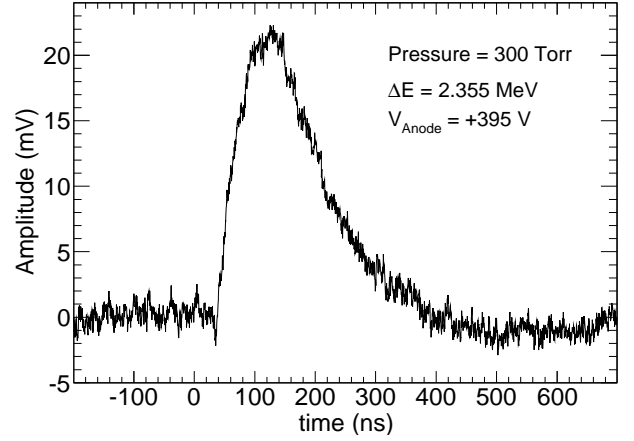}
\caption{
Typical signal from the CSA corresponding to the energy loss of an $^{241}$Am $\alpha$ particle with 300 torr of CF$_{4}$ gas in the detector.}
\label{fig:RIPD_CSA_Signal}
\end{center}
\end{figure}

A typical signal from the CSA at a pressure of 300 torr is shown in Fig.~\ref{fig:RIPD_CSA_Signal}. 
The anode was biased to a potential of +395 V to produce a reduced field of 1 kV cm$^{-1}$ atm$^{-1}$.
The signal time from baseline to peak is approximately 100 ns with a rise time of 60 ns. 
Although the observed signal rise time corresponds to 
the convolution of the electron collection time and the CSA response, 
as the CSA response is fast ($<$10 ns), the observed signal 
principally reflects the electron collection time, consistent 
with the reported literature value for the electron drift velocity at the reduced field utilized.
The CSA signal returns 
to baseline after 300 ns. Thus the whole signal duration is under 500 ns, 
which corresponds to a maximum calculated rate of 
$\sim$2$\times$10$^{6}$ ions/second without pileup. The gain of the CSA is approximately 9.5 mV/MeV. 
The noise of this signal is approximately 4 mV peak-to-peak, corresponding to a signal-to-noise ratio of 5.7.

To handle these fast signals, the development of a fast shaping amplifier was required. 
This requirement was realized by modifying  an in-house octal shaper module 
which handles input signals of both polarities. 
With the fast shaper module shaping times between 
100 ns and 800 ns in increments of 100 ns can be selected. The coarse gain is controlled by two 4-bit stages, 
while a fine gain adjustment is provided using a 12-bit multiplying ADC. 
Another 12-bit multiplying ADC allows adjustment of the pole-zero.
All of these parameters can be adjusted under computer control through a USB 2.0 interface. 
With a shaping time of 200 ns the fast shaping amplifier
transforms the typical input CSA signal depicted in Fig.\ref{fig:RIPD_CSA_Signal}
into a
Gaussian-like pulse shown with an amplitude of $\sim$950 mV and a peak-to-peak high 
frequency noise of $\sim$30 mV. 
Thus the shaping amplifier improves the signal-to-noise ratio to a value of approximately 31.

\section{Energy Resolution}

\begin{figure}
\begin{center}
\includegraphics[scale=0.44]{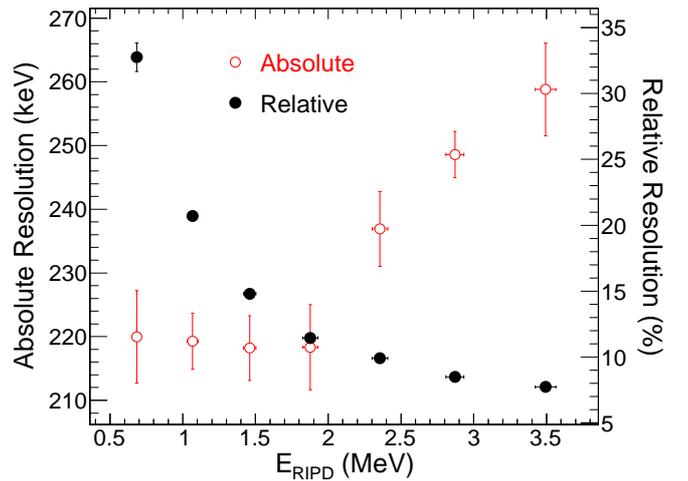}
\caption{
Resolution as a function of energy deposit in the gas volume of RIPD
for $\alpha$ particles from a $^{241}$Am source.}
\label{fig:RIPD_ResPlot}
\end{center}
\end{figure}

\begin{figure*}
\begin{center}
\includegraphics[scale=0.44]{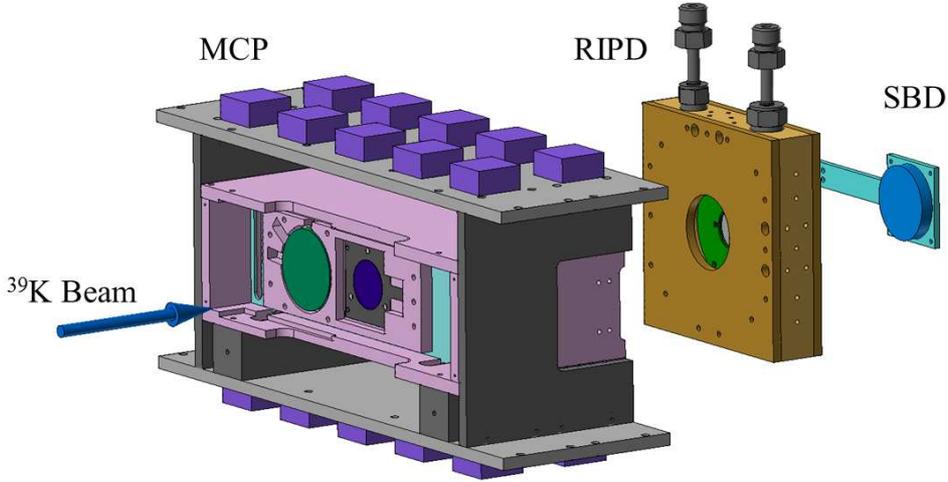}
\caption{
CAD representation of the experimental setup used to characterize RIPD 
with beam.}
\label{fig:beam_setup}
\end{center}
\end{figure*}

To determine the energy resolution of RIPD, the signal from the 
shaping amplifier was digitized by a CAEN V785 peak-sensing ADC. 
The energies of particles measured in 
RIPD was first calibrated using the measured energies in the SBD detector after accounting for the
energy lost in the mylar foils \cite{SRIM}. The calibrated RIPD energy spectrum was used to determine the energy resolution by comparing the FWHM of the 
energy distribution to its centroid. The dependence of the 
energy resolution on
the energy deposited in the gas volume 
is shown in Fig.~\ref{fig:RIPD_ResPlot}. 
For energies up to $\sim$2 MeV, 
the absolute resolution is constant at a value of approximately 220 keV, 
suggesting that for these energy deposits the electronic noise dominates the total noise. Above 2 MeV the 
absolute resolution deteriorates reaching a value of $\sim$260 keV at 3.5 MeV of energy deposit.
In the energy deposit interval measured the relative resolution decreases smoothly from 32\% to $\sim$7.5\%.
This 7.5\% resolution can be understood as a combination of the signal-to-noise 
after the fast shaper as well as the variations in the path 
length of the measured $\alpha$ particles through the gas volume.

\section{Performance with Beam}
In order to characterize the performance of RIPD with beams of different intensity a test was conducted at 
the ReA3 facility at
Michigan State University's National Superconducting Cyclotron Laboratory (NSCL). The ReA3 80 MHz linac, 
which can be used to accelerate either stable beams or 
radioactive ions produced by the NSCL's coupled cyclotron facility, was used to accelerate $^{39}$K to 
energies of 2.5 MeV/A and 4 MeV/A. The beam was 
extracted from the charge breeding ion trap EBIT within 100 milliseconds at a repetition rate of 2 Hz. 
This time structure results in the 
instantaneous rate experienced by any detector in the beam path being effectively a factor of five higher 
than the average rate. 
The experimental setup used is depicted in Fig.~\ref{fig:beam_setup}. The first element of the setup was 
a microchannel plate 
detector \cite{Steinbach14a}.
In this detector, passage of a beam particle through a 100 $\mu$g/cm$^2$ carbon foil ejects electrons 
that are transported by crossed electric and magnetic fields 
to the surface of a chevron microchannel plate (MCP). The fast response of the MCP results in each 
beam ion incident on the carbon foil being 
individually recorded. The rate at which the MCP triggers was recorded by a  250 MHz VME scaler providing a measure of the beam rate. 
Approximately 44 cm downstream of the MCP, 
RIPD was mounted on a retractable arm. For low intensity beams a SBD 
situated immediately after RIPD was used to measure the residual energy of ions traversing RIPD. 
The SBD was retracted from the beam path when a high intensity beam was used.

\begin{figure}
\begin{center}
\includegraphics[scale=0.44]{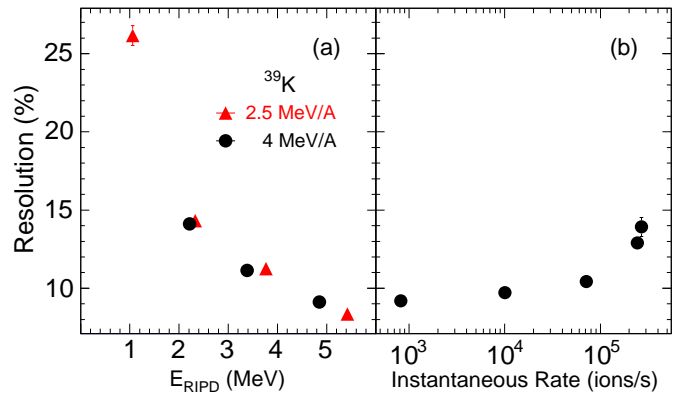}
\caption{Panel a: Dependence of the energy resolution on the deposited energy in RIPD. Panel b: Impact on energy resolution
of increased instantaneous beam rate.}
\label{fig:beam_resolution}
\end{center}
\end{figure}

Shown in Fig.~\ref{fig:beam_resolution}a as the closed triangles is the energy resolution of the energy deposited in RIPD by 
2.5 MeV/A $^{39}$K ions 
traversing RIPD at nominal gas pressures of
5, 10, 15, and 20 torr. The gas pressure and flow through the detector were maintained by a high quality 
gas handling system with a
stability of 0.2 torr. With increasing gas pressure the energy deposit in the gas increases from 
approximately 1 MeV to 
approximately 5 MeV. As the energy deposit in the gas increases the relative resolution of RIPD improves 
from 25$\%$ at the 
lowest energy deposit to approximately 9$\%$ at the highest energy deposit in 
reasonable agreement with the source testing results presented in Fig.~\ref{fig:RIPD_ResPlot}.  
The electronic noise as well as 
the stability of the electronics were monitored
during the measurement by injecting a calibration pulse into the charge sensitive amplifier at a low rate. The electronic noise was measured to be 
approximately 1 $\%$, significantly less than the measured energy resolution of the detector.
To determine the response of the energy resolution of RIPD to a
change in the beam intensity, the beam intensity of a 4 MeV/A $^{39}$K beam was increased from 800 ions/s to 
3 x 10$^5$ ions/s. 
The impact of the beam intensity on the energy resolution of RIPD is shown in Fig.~\ref{fig:beam_resolution}b. 
With increasing 
beam intensity the measured energy resolution degrades from approximately 
9$\%$ to $\sim$14$\%$ at the highest intensity measured.

\begin{figure}
\begin{center}
\includegraphics[scale=0.44]{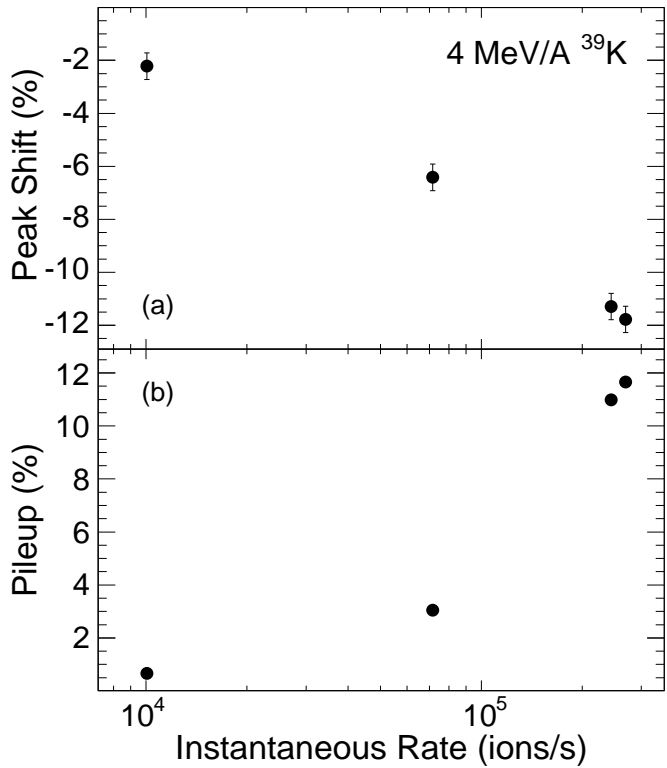}
\caption{
Impact of the beam intensity on the charge collection (panel a) and pileup (panel b) in RIPD.}
\label{fig:beam_shiftpileup}
\end{center}
\end{figure}

Aside from the degradation in the energy resolution the increased beam intensity can also impact the 
total charge collection due to increased recombination in the detector at high beam intensities. This effect 
is manifested in the upper panel of Fig.~\ref{fig:beam_shiftpileup}. 
A decrease in the centroid of the energy deposited in RIPD by 4 MeV/A $^{39}$K ions
is evident.
With increasing beam intensity 
the location of the energy centroid shifts to lower values consistent with recombination of electrons and cations
in the detector gas. At an instantaneous rate of 3 x 10$^5$ ions/s a peak shift of $\sim$12\% is observed 
as compared to a low intensity beam of 800 ions/s. As recombination impacts both the 
ions of interest as 
well as any contaminants, identifying the nuclide of primary interest from other contaminants is 
still achieveable. However, correcting for any significant changes in the beam intensity becomes important.

The impact of beam intensity on the pileup observed is indicated in the lower panel 
of Fig.~\ref{fig:beam_shiftpileup}. Pileup is clearly distinguished 
as observation of 
pulses with energies exceeding that of the full beam energy. 
As expected the pileup increases with increasing instantaneous rate from 
$<$1\% at 1 x 10$^4$ ions/s to just under 12\% at 3 x 10$^5$ ions/s. As anticipated the fast 
response of RIPD enables the use of the detector at high rates without significant pileup.

\section{Conclusion}

To address the challenge of identifying contaminant species in radioactive isotope beams, 
we have developed a high-rate axial-field ionization chamber which introduces minimal material into the beam 
path. The high-rate capability of 
this detector was optimized by implementing a low profile central anode design and 
developing a fast charge-sensitive amplifier which is situated within the gas volume 
in close proximity to the anode. A typical signal from this amplifier 
spans less than 500 ns, with a rise time of approximately 60-70 ns. A fast shaping amplifier was also 
developed to handle these fast signals. With these pulse shaping 
electronics the energy resolution of RIPD is
approximately 8\% for an energy deposit of 3.5 MeV in the detector for $\alpha$ particles from an 
$^{241}$Am source. Using beam, a comparable resolution was obtained. Below an instantaneous rate of 1 x 10$^5$ 
$^{39}$K ions/s the energy resolution was 8-10\%. At an instantaneous rate of 3 x 10$^5$ the resolution degraded 
to 14\%. The impact of recombination and pileup in the detector as a function of the instantaneous rate was 
characterized. Correcting for these effects is an important part of exploiting the full capability 
of this detector to resolve contaminants in a radioactive beam.

\section*{Acknowledgments}
We gratefully acknowledge the technical support provided by the personnel in the 
Mechanical Instrument Services and Electronic Instrument Services at the Department of Chemistry,
Indiana University. We thank the technical staff at MSU-NSCL for providing the high quality beam necessary to fully characterize the detector.
This research is based upon work supported by the U.S. Department of Energy under Award Number DE-FG02-88ER-40404 
and the National Science Foundation under Grant Number 1342962.
 
%\bibliographystyle{elsarticle-num}
%\bibliography{RIPD_NIM}

\end{document}